\documentclass[preprint,aps,prd,showpacs,superscriptaddress,nofootinbib,tightenlines]{revtex4-1}

\usepackage{amsfonts}
\usepackage{multirow}
\usepackage{mathrsfs}
\usepackage{graphicx}
\usepackage{amsmath}
\usepackage{amssymb}
\usepackage{bm}
\usepackage{color}
\usepackage{hyperref}
\usepackage{slashed}

\usepackage{ulem} 

\newcommand{\nn}{\nonumber}

\newcommand{\beq}{\begin{equation}}
\newcommand{\eeq}{\end{equation}}

\newcommand{\bseq}{\begin{subequations}}
\newcommand{\eseq}{\end{subequations}}

\begin{document}


\title{\mbox{}\\[14pt]
Four-lepton decays of neutral vector mesons
}

\author{Wen Chen~\footnote{wchen1@ualberta.ca}}
\affiliation{Institute of High Energy Physics, Chinese Academy of
Sciences, Beijing 100049, China\vspace{0.2 cm}}
\affiliation{Department of Physics, University of Alberta, Edmonton, AB T6G 2E1, Canada\vspace{0.2 cm}}

\author{Yu Jia~\footnote{jiay@ihep.ac.cn}}
\affiliation{Institute of High Energy Physics, Chinese Academy of
Sciences, Beijing 100049, China\vspace{0.2 cm}}
\affiliation{School of Physics, University of Chinese Academy of Sciences,
Beijing 100049, China\vspace{0.2 cm}}

\author{Zhewen Mo~\footnote{mozw@ihep.ac.cn}}
\affiliation{Institute of High Energy Physics, Chinese Academy of
	Sciences, Beijing 100049, China\vspace{0.2 cm}}
\affiliation{School of Physics, University of Chinese Academy of Sciences,
	Beijing 100049, China\vspace{0.2 cm}}

\author{Jichen Pan~\footnote{panjichen@ihep.ac.cn}}
\affiliation{Institute of High Energy Physics, Chinese Academy of
	Sciences, Beijing 100049, China\vspace{0.2 cm}}
\affiliation{School of Physics, University of Chinese Academy of Sciences,
	Beijing 100049, China\vspace{0.2 cm}}

\author{Xiaonu Xiong~\footnote{xnxiong@csu.edu.cn}}
\affiliation{School of Physics and Electronics, Central South University, Changsha 418003, China}

\date{\today}

\begin{abstract}
We investigate the rare electromagnetic decays of neutral vector mesons (exemplified by $J/\psi$, $\Upsilon$, $\rho^0$,
$\omega,\phi$, \ldots) into four charged leptons ($2(e^+e^-)$, $2(\mu^+\mu^-)$, $2(\tau^+\tau^-)$,
$e^+e^-\mu^+\mu^-$, $e^+e^-\tau^+\tau^-$, \ldots) at lowest order in QED.
In contrast to the case of vector meson decay into a single pair of leptons,
the lepton mass must be retained in these four-lepton decay channels to cutoff the mass singularity.
Owing to more pronounced collinear enhancement, the branching fractions of vector mesons decays into $ 2(e^+e^-)$ are
considerably greater than those for decays into $2(\mu^+\mu^-)$.
The decay channels $J/\psi(\Upsilon)\to 2(e^+e^-), e^+e^-\mu^+\mu^-$ are
predicted to have branching fractions of order $10^{-5}$,
which appear to have bright observation prospect at {\tt BESIII}, {\tt Belle 2} and {\tt LHC} experiments.
\end{abstract}

\maketitle

The leptonic decays of neutral vector mesons are arguably one of the cleanest and most exhaustively studied hadronic processes.
To dematerialize into a lepton pair, the valence quark and antiquark inside the vector meson have to first
annihilate into a virtual photon, which then subsequently transition into a lepton pair.
Since the hadronic decay part and the electromagnetic production part of the amplitude
simply factorizes, one can write down the well-known formula for vector meson leptonic decay:
\beq
\Gamma\left(V\rightarrow e^{+}e^{-}\right) = \frac{4\pi e_{q}^{2}\alpha^{2}f_{V}^{2}}{3M_{V}},
\label{LO:NRQCD:Jpsi:ee}
\eeq
where $\alpha$ denotes the fine structure constant,
$e_q$ signifies the electric charge of the valence quark of $V$.
$M_V$ denotes the vector meson mass, and $f_V$ denotes the decay constant associated with $V$~\footnote{If $V$ indicates a heavy quarkonium,
the decay constant can then be approximated by $f_{V} \approx \sqrt{\frac{3}{\pi M_{V}}}R_{V}(0)$, with
the factor $R_V(0)$ signifying the radial wave function at the origin for a $^3S_1$ quarkonium.}:
\beq
\langle0\vert\bar{q}\gamma^{\mu}q\vert V\left(P\right)\rangle = f_{V}M_{V}\varepsilon^{\mu}\left(P\right),
\label{decay:constant}
\eeq
where $\varepsilon^\mu$ signifies the polarization vector of the $V$ meson.

In \eqref{LO:NRQCD:Jpsi:ee}, the lepton mass has been neglected for simplicity. It is obviously a perfect
approximation for the vector meson decay into electrons, and still a decent one for heavy vector quarkonia such as
$J/\psi$ and $\Upsilon$ decay into a muon pair.

\begin{table*}[!h]
  \begin{tabular}{|c|c|c|c|}
    \hline
     & $\rho^{0}$ & $\omega$ & $\phi$\tabularnewline
    \hline
    $e^{+}e^{-}$ & $\left(4.72\pm0.05\right)\times10^{-5}$ & $\left(7.36\pm0.15\right)\times10^{-5}$ & $\left(2.973\pm0.034\right)\times10^{-4}$\tabularnewline
    \hline
    $\mu^{+}\mu^{-}$ & $\left(4.55\pm0.28\right)\times10^{-5}$ & $\left(7.4\pm1.8\right)\times10^{-5}$ & $\left(2.86\pm0.19\right)\times10^{-4}$\tabularnewline
    \hline
    \end{tabular}
\caption{The measured ${\cal B}(V \to 2l)$ for $V=\rho^0,\omega,\phi$ mesons~\cite{pdg:2020}.}
\label{T1}
\end{table*}

\begin{table*}[!h]
  \begin{tabular}{|c|c|c|}
    \hline
     & $J/\psi$ & $\Upsilon$\tabularnewline
    \hline
    $e^{+}e^{-}$ & $\left(5.971\pm0.032\right)\%$ & $\left(2.38\pm0.11\right)\%$\tabularnewline
    \hline
    $\mu^{+}\mu^{-}$ & $\left(5.961\pm0.033\right)\%$ & $\left(2.48\pm0.05\right)\%$\tabularnewline
    \hline
    $\tau^{+}\tau^{-}$ & $/$ & $\left(2.60\pm0.10\right)\%$\tabularnewline
    \hline
    \end{tabular}
\caption{The measured ${\cal B}(V \to 2l)$ for $V=J/\psi, \Upsilon$~\cite{pdg:2020}.}
\label{T2}
\end{table*}

As one can tell from Table~\ref{T1} and ~\ref{T2},
the branching fractions of various leptonic decays of vector mesons
have been measured quite precisely. From these experimental inputs,
one sees that discarding the lepton mass is indeed a satisfactory approximation in
\eqref{LO:NRQCD:Jpsi:ee}, and one can also deduce the values of the nonperturbative factor $f_V$ for
different vector mesons.

At present, {\tt BESIII}~\cite{Asner:2008nq}, {\tt Belle 2}~\cite{Jia:2020csg}, {\tt KLOE} experiment~\cite{KLOE:2004uzx}, and
{\tt LHC} experiments~\cite{LHCb:2013itw}
have accumulated a gigantic number of $J/\psi$ and $\Upsilon$,
as well as light vector mesons $\rho$, $\omega$ and $\phi$, which make it feasible to search some truly rare decay channels.
The aim of this paper is to explore a special kind of rare electromagnetic
decay processes, that is, neutral vector meson decays into four charged leptons.
Concretely speaking, we are interested in considering two types of electromagnetic
decays, $V\to 2(l^+l^-)$ and $l^+l^- l^{\prime+}l^{\prime-}$ with $l,l^\prime=e,\mu,\tau$.
Such multi-lepton rare decay of quarkonia represents a novel testing bed for QED predictions, and
we hope they can be seen at {\tt BESIII}, {\tt Belle 2}, {\tt KLOE} and {\tt LHC} experiments in near future.

The partial width of neutral vector meson decay to four leptons reads
\begin{eqnarray}
&& \Gamma(V \to 4l) = {1\over 2M_V} \int \!\! d\Pi_4 {1\over 3} \sum_{\text{spins}}\left|{\cal M}(V\to 4l)\right|^2
=\frac{e_q^2e^2f_V^2}{6M_V^3}\int\!\! d\Pi_4 \sum_{\mathrm{spins}} \left|\mathcal{M}(\gamma^*\rightarrow 4l)\right|^2,
\nn
\\
\label{Gamma:V:4l}
\end{eqnarray}
where $d\Pi_4$ is the four-body phase space with possible symmetry factors due to identical leptons.
We will dedicate Appendix~\ref{App:FourBodPhasSpacInt} for a detailed account of
the technicalities related to 4-body phase space integration.
In the second equality, we utilize the fact that, entirely analogous to the process $V\to 2l$,
the decay amplitude simply factorizes into the product of the
$V\to \gamma^*$ hadronic part and the $\gamma^*\to 4l$ leptonic part.

\begin{figure}[hbtp]
\centering
\includegraphics[width=1.0\textwidth]{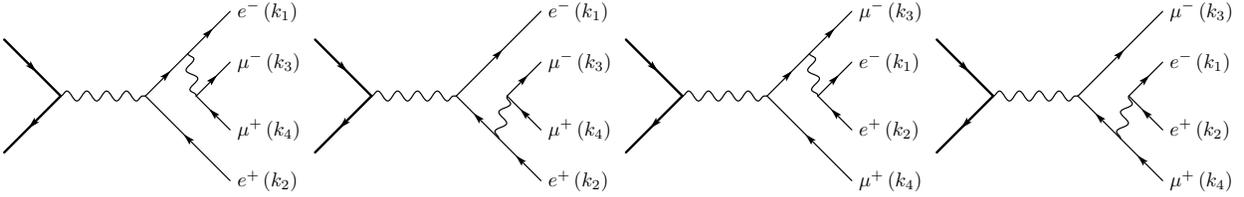}
\caption{Feynman diagrams for $V\to \gamma^*\to e^+e^-\mu^+\mu^-$. }
\label{Fig_eemumu}
\end{figure}

\begin{figure}[hbtp]
  \centering
  \includegraphics[width=0.85\textwidth]{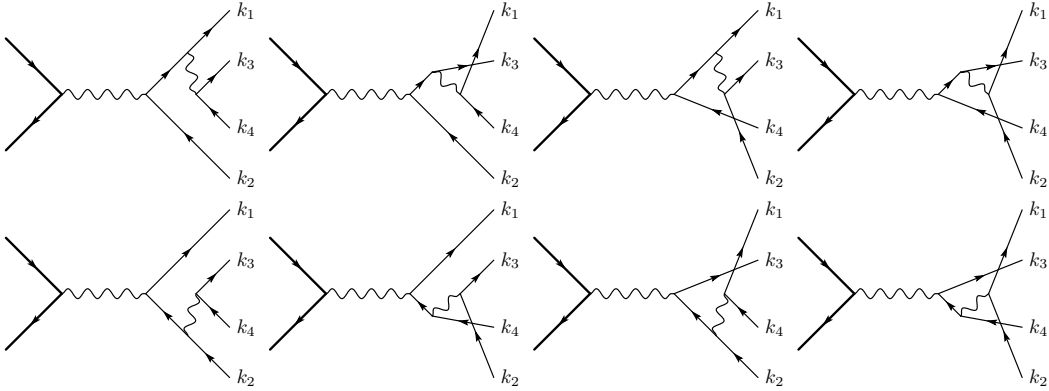}
  \caption{Feynman diagrams for $V\to \gamma^*\to 2( e^+e^-)$.}
  \label{Fig_eeee}
\end{figure}

We employ the packages {\tt FeynArts}~\cite{Hahn:2000kx} to generate the Feynman diagrams for $V\to 4l$.
The corresponding four Feynman diagrams for $V\to l^+l^- l^{\prime+}l^{\prime-}$ have been displayed in
Fig.~\ref{Fig_eemumu}, while the corresponding eight diagrams for $V\to 2(l^+l^-)$
$V \to 2(e^+e^-)$ are illustrated in Fig.~\ref{Fig_eeee}.
Subsequently we use {\tt FeynCalc}~\cite{Shtabovenko:2020gxv} to handle
Dirac trace/Lorentz tensor algebra.

For phenomenological purpose, it is more convenient to introduce the following dimensionless ratio $R$,
rather than directly compute the partial width:
\beq
R(V \to 4l) \equiv \frac{\Gamma(V\to 4l)}{\Gamma(V\to e^+e^-)}.
\label{Def:R:ratio}
\eeq
As is evident in \eqref{LO:NRQCD:Jpsi:ee} and \eqref{Gamma:V:4l}, the advantage of introducing the $R$ ratio
is that the nonpertubative factor $f_V$, as well as the electric charge of valence quark $e_q$,
exactly cancel between the numerator and the denominator.
Therefore, the prediction of the $R$ ratio is dictated entirely by QED,
without contamination by any nonperturbative effects.
Since $R$ is dimensionless, it can only be a function of the mass ratio $m_l/M_V$.
We envisage that the leading order QED prediction should suffice to
provide a decent account for the $R$ values for various $V\to 4l$ channels.

In \eqref{LO:NRQCD:Jpsi:ee} we have safely discarded the lepton mass in $V \to l^+l^-$ since $m_{e,\mu}\ll M_V$.
One may wonder whether the same simplification can be taken or not for $V \to 4l$.
As a matter of fact, for the four-body decays, it is possible that three particles can move collinear and some are
simultaneously soft, in which the collinear and soft singularity may be developed.
In fact, to assess the degreee of mass singularity, we invoke the reverse unitarity
method~\cite{Anastasiou:2002yz} to analytically compute the $R$ ratio for vector meson $V$ decay to four massless leptons (as depicted in
Fig.~\ref{Fig_eeee}) in spacetime dimension $d=4-2\epsilon$:
\begin{align}
R(V \to 4l) = & {1\over \Gamma^3(1-\epsilon)} \left(\frac{M_V^2}{4\pi\mu^2}\right)^{-3 \epsilon }\left(\frac{\alpha}{2\pi}\right)^2\left[-\frac{1}{3 \epsilon ^3}-\frac{17}{9 \epsilon ^2}+\frac{1}{\epsilon }\left(-2 \zeta (3)+\frac{10 \pi ^2}{9}-\frac{1405}{108}\right)\right.
\nn \\
&~ \left.+\frac{103 \zeta (3)}{3}+\frac{367 \pi ^2}{54}-\frac{8 \pi ^4}{45}-\frac{60835}{648}\right],
\label{NNLO:massless:R}
\end{align}
where $\mu$ is the 't Hooft unit mass. Note that after supplemented with proper color factor,
this piece may be identified with the Abelian subset of the double-real part
of the next-to-next-leading order (NNLO) QCD corrections for
$e^+e^-\to \gamma^* \to q\bar{q}$, which has already been known analytically long ago~\cite{Gehrmann-DeRidder:2003pne,GehrmannDeRidder:2004tv}.
Note the occurrence of the triple infrared pole in \eqref{NNLO:massless:R} unambiguously
indicates that the $R$ ratio in \eqref{Def:R:ratio} does not have a safe $m_l\to 0$ limit, in sharp contrast to
$V \to e^+e^-$.

In our numerical study we work in $d=4$ spacetime dimension and keep $m_l$ finite to regularize the emerging mass singularity.
It is intuitively envisaged the leading $1/\epsilon^3$ pole in \eqref{NNLO:massless:R} would be literally translated into
the triple-logarithm $\ln^3(M_V^2/m_l^2)$ in our case.
Expanding the $\mu$-dependent prefactor  in \eqref{NNLO:massless:R} to order $\epsilon^3$,
multiplying with the $1/\epsilon^3$ pole,  replacing $\mu$ by $m_l$,
we readily identify the leading triple-logarithmic term:
\begin{equation}
R(V \to 4l) \approx  \frac{3\alpha^2}{8\pi^2}\ln^3 {M_V^2\over m_l^2}+
{\cal O}\left(\ln^2\frac{M_V^2}{m_l^2}\right).
\label{triple:logarithm}
\end{equation}

It is illuminating to trace the origin of the  triple-logarithm in \eqref{triple:logarithm} for the case of
realistic lepton, or equivalently,
the $1/\epsilon^3$ pole in \eqref{NNLO:massless:R} for the case of massless lepton.
Near the tri-collinear limit, {\it i.e.}, when one of the $l^+$ and two $l^-$ become nearly collinear,
the squared amplitude for $\gamma^*\to 4l$ can be factorized into the squared amplitude for
$\gamma^*\to l^+l^-$ times the universal NNLO splitting function for
$l^-\to l^-+(l^-\l^+)$~\cite{Catani:1999ss}.
Double collinear poles would immediately arise in the tri-collinear limit.
Furthermore, when the invariant mass of the $l^-l^+$-pair also becomes simultaneously soft, the adjacent photon propagator
also becomes soft, which would develop an additional soft pole.
We have explicitly verified that, upon integrating the NNLO
splitting function over three-lepton phase space, the leading triple IR pole in
\eqref{NNLO:massless:R} can indeed be recovered~\cite{Magnea:2018hab}.

For numerical analysis, we adopt the following values for various input parameters~\cite{pdg:2020}:
\begin{align}
& \alpha = 1/137, \quad m_e = 0.511\;\text{MeV}, \quad m_\mu = 0.106\;\text{GeV}, \quad m_\tau = 1.78\;\text{GeV},\nn\\
& M_\rho = 0.77526\;\text{GeV}, \quad M_\omega = 0.78265\;\text{GeV} \quad M_\phi = 1.019461\;\text{GeV},\nn\\
& M_{J/\psi} = 3.0969\;\text{GeV}, \quad M_\Upsilon = 9.4603\;\text{GeV}.
\end{align}

We use the numerical package {\tt CUBA}~\cite{Hahn:2016ktb} to conduct the 4-body phase space integration,
which implements the adaptive Monte Carlo integration algorithm.

\begin{figure}[hbtp]
\centering
\includegraphics[width=.6\linewidth]{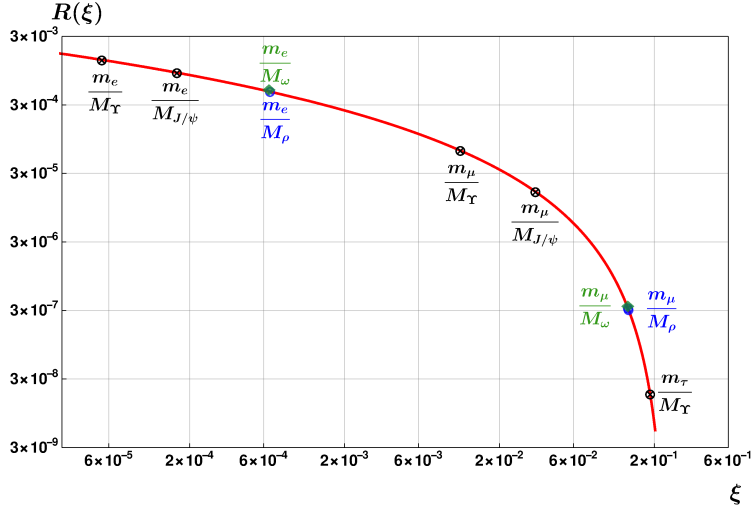}
\caption{
The ratio $R(V\to 4l)$ as a function of $\xi=\frac{m_l}{M_V}$.
Note that this is a log-log plot, and the logarithmical divergence is reflected by the negative slope of the curve
in the $\xi \to 0$ limit.}
\label{Fig:R:as:function:of:xi}
\end{figure}

The dependence of the $R$ ratio for $V\to 4l$ on $\xi\equiv m_l/M_V$ is shown in
Fig.~\ref{Fig:R:as:function:of:xi}. It can be readily visualized that the curve becomes singular as $\xi\to 0$.
A numerical fit reveals that the small-$m_l$ enhancement is compatible with the
triply-logarithmical scaling behavior.

\begin{table*}
  \resizebox{\textwidth}{!}{
  \begin{tabular}{|c|c|c|c|c|c|c|}
    \hline
     & \multicolumn{3}{c|}{$R\left(\times10^{-4}\right)$} & \multicolumn{3}{c|}{$\mathcal{B}\left(\times10^{-8}\right)$}\tabularnewline
    \hline
     & $\rho^{0}$ & $\omega$ & $\phi$ & $\rho^{0}$ & $\omega$ & $\phi$\tabularnewline
    \hline
    $e^{+}e^{-}\mu^{+}\mu^{-}$ & $2.05$ & $2.08$ & $2.74$ & $0.970\pm0.010$ & $1.528\pm0.031$ & $8.14\pm0.09$\tabularnewline
    \hline
    $2\left(e^{+}e^{-}\right)$ & $4.68$ & $4.71$ & $5.37$ & $2.211\pm0.023$ & $3.46\pm0.07$ & $15.97\pm0.18$\tabularnewline
    \hline
    $2\left(\mu^{+}\mu^{-}\right)$ & $3.09\!\times\!10^{-3}$ & $3.26\!\times\!10^{-3}$ & $0.0114$ & $\left(1.456\!\pm\!0.015\right)\!\times\!10^{-3}$ & $\left(2.400\!\pm\!0.049\right)\!\times\!10^{-3}$ & $\left(3.383\!\pm\!0.039\right)\!\times\!10^{-2}$\tabularnewline
    \hline
    \end{tabular}
  }
    \caption{Predictions for $R$ ratios and the affiliated branching fractions for $\text{light vector mesons} \to 4l$. Note the latter is obtained
    by multiplying $R$ with the corresponding measured branching fraction of $V\to l^+l^-$, with our theoretical
    uncertainty entirely inherited from experimental error on di-leptonic decay of $V$.}
\label{Tab:R:ratio:Branching:ratio:Light:mesons}
\end{table*}

\begin{table*}
  \begin{tabular}{|c|c|c|c|c|}
    \hline
     & \multicolumn{2}{c|}{$R\left(\times10^{-4}\right)$} & \multicolumn{2}{c|}{$\mathcal{B}\left(\times10^{-5}\right)$}\tabularnewline
    \hline
     & $J/\psi$ & $\Upsilon$ & $J/\psi$ & $\Upsilon$\tabularnewline
    \hline
    $e^{+}e^{-}\mu^{+}\mu^{-}$ & $6.31$ & $11.5$ & $3.763\!\pm\!0.020$ & $2.73\!\pm\!0.13$\tabularnewline
    \hline
    $e^{+}e^{-}\tau^{+}\tau^{-}$ & $/$ & $3.74$ & $/$ & $0.890\!\pm\!0.041$\tabularnewline
    \hline
    $\mu^{+}\mu^{-}\tau^{+}\tau^{-}$ & $/$ & $0.252$ & $/$ & $0.0600\!\pm\!0.0028$\tabularnewline
    \hline
    $2\left(e^{+}e^{-}\right)$ & $8.85$ & $13.7$ & $5.288\!\pm\!0.028$ & $3.25\!\pm\!0.15$\tabularnewline
    \hline
    $2\left(\mu^{+}\mu^{-}\right)$ & $0.163$ & $0.648$ & $0.0974\!\pm\!0.0005$ & $0.154\!\pm\!0.007$\tabularnewline
    \hline
    $2\left(\tau^{+}\tau^{-}\right)$ & $/$ & $1.82\!\times\!10^{-4}$ & $/$ & $\left(4.33\pm0.20\right)\times10^{-5}$\tabularnewline
    \hline
    \end{tabular}
    \caption{Predictions for $R$ ratios and the affiliated branching fractions for $J/\psi,\Upsilon \to 4l$. Note the latter is obtained
    by multiplying $R$ with the measured branching fraction for $V\to l^+l^-$, with our theoretical uncertainty entirely
    propagating from experimental error. }
    \label{Tab:R:ratio:Branching:ratio:Quarkonium}
  \end{table*}

Our detailed predictions for $\rho^0,\omega,\phi \to 4l$ and $J/\psi,\Upsilon\to 4l$ are tabulated in Table~\ref{Tab:R:ratio:Branching:ratio:Light:mesons} and
Table~\ref{Tab:R:ratio:Branching:ratio:Quarkonium}.
By multiplying the measured branching fractions for $V\to l^+l^-$ as given in Table~\ref{T1} and ~\ref{T2},
we then predict the branching fractions for $V\to 4l$.
As one can see, the branching fractions for $\rho^0,\omega \to 2(e^+e^-), e^+e^-\mu^+\mu^-$ are of order $10^{-8}$,
while those for $\phi \to 2(e^+e^-), e^+e^-\mu^+\mu^-$ may reach $10^{-7}$.
The rather tiny branching fractions make the observation of these rare decays of light vector mesons a great challenge.
On the other hand, the branching fractions of $J/\psi \to 2(e^+e^-), e^+e^-\mu^+\mu^-$ can reach the order of $10^{-5}$~\footnote{Needless
to emphasize, it should be obvious that all our predictions for the $R$ ratios in Table~\ref{Tab:R:ratio:Branching:ratio:Quarkonium}
can be carried over to the four-lepton decays of excited vector quarkonia such as $\psi(2S)$ and $\Upsilon(2S,3S)$.}.
Concerning the billions of $J/\psi$ events collected at {\tt BESIII}, it is hopeful to discover these two channels
in near future. In contrast, because muon is much heavier than electron, the triple logarithmic enhancement for
$J/\psi \to 2(\mu^+\mu^-)$ is much less pronounced. From Table~\ref{Tab:R:ratio:Branching:ratio:Quarkonium}, we observe
${\cal B}(J/\psi \to 2(\mu^+\mu^-))$ is only about $2$\% of  ${\cal B}(J/\psi \to 2(e^+e^-))$, which renders the observation
prospect in near future obscure.
The pattern of predictions for $\Upsilon\to 4l$ is similar. ${\cal B}(\Upsilon \to 2(e^+e^-)),  e^+e^-\mu^+\mu^-), e^+e^-\tau^+\tau^-)$
are of order $10^{-5}$, while ${\cal B}(\Upsilon \to 2(\mu^+\mu^-)), \mu^+\mu^-\tau^+\tau^-)$ are $1 \sim 2$ orders of magnitude smaller.
Compared with these channels, ${\cal B}(\Upsilon \to 2(\tau^+\tau^-))$ is too tiny to have any chance for observation.
It is interesting to note that, the branching fractions associated with three decay channels $\Upsilon \to 2(e^+e^-),
e^+e^-\mu^+\mu^-$ and $\Upsilon \to e^+e^-\tau^+\tau^-$ do not exhibit considerable hierarchy, all of which is of order $10^{-5}$.
We hope the dedicated analysis in {\tt Belle 2} and perhaps {\tt LHC} will observe these channels in the foreseeable future.

\begin{figure}[hbtp]
	\centering
  \includegraphics[width=0.45\textwidth]{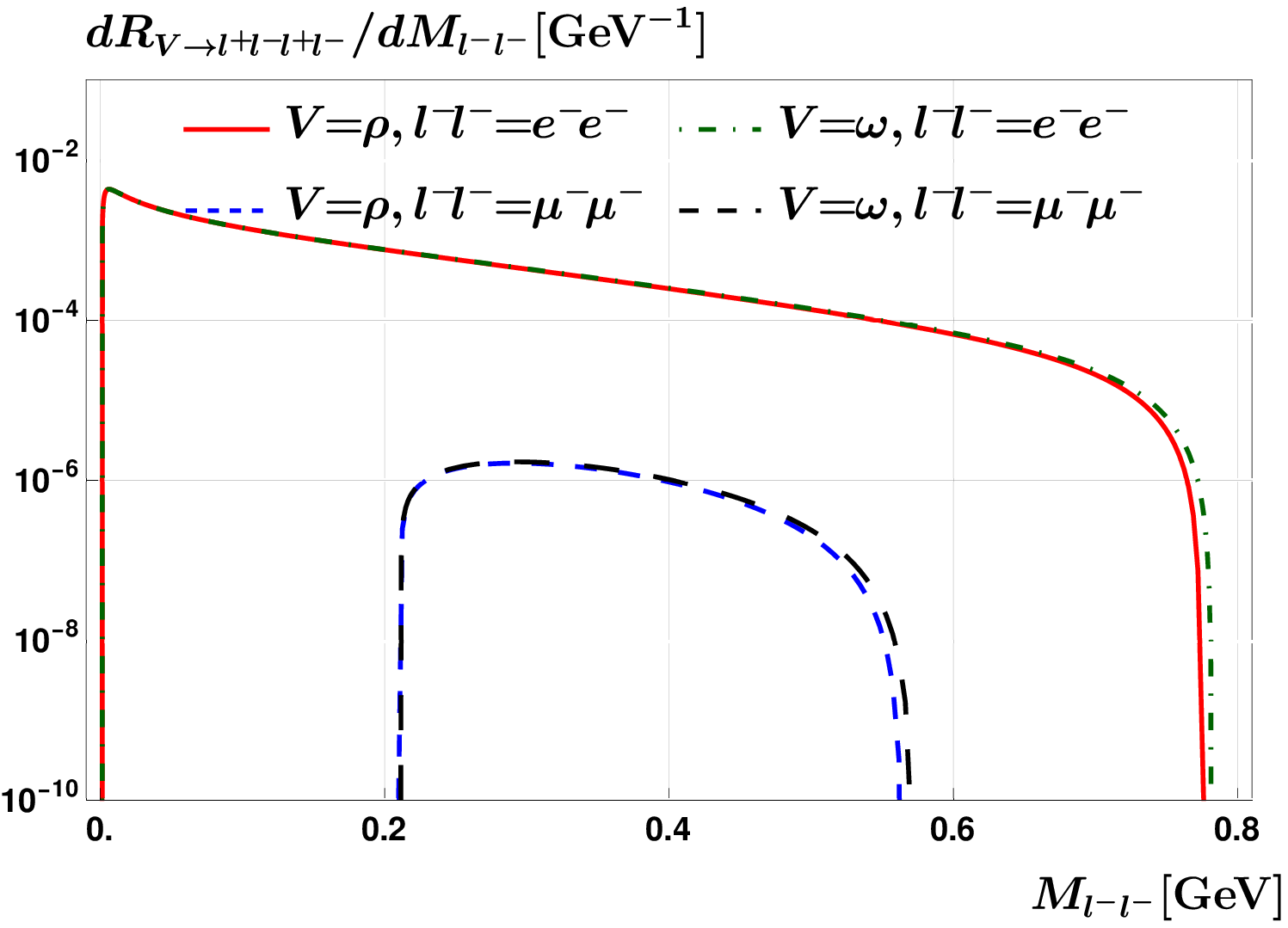}
  \includegraphics[width=0.45\textwidth]{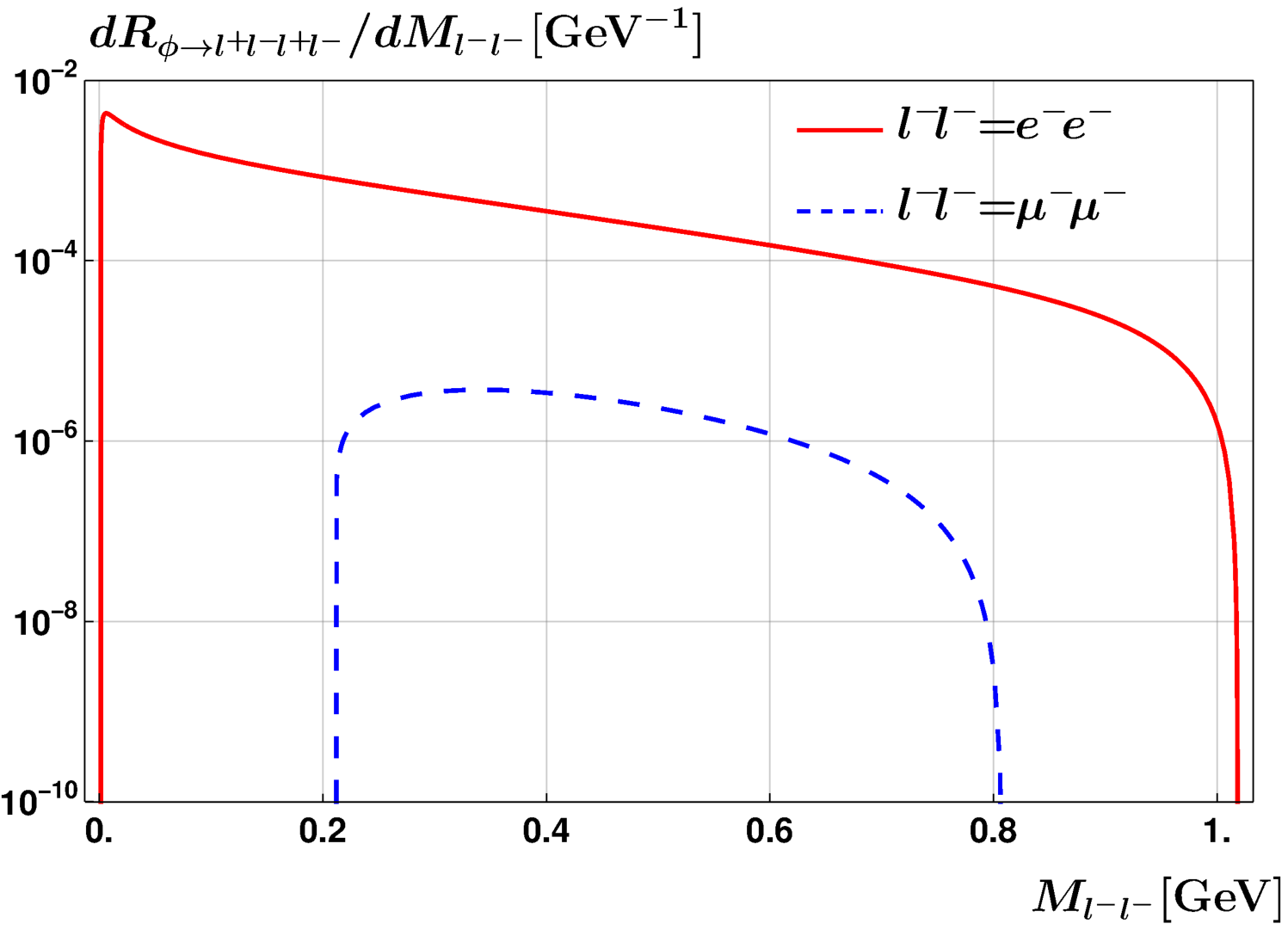}
	\includegraphics[width=0.45\textwidth]{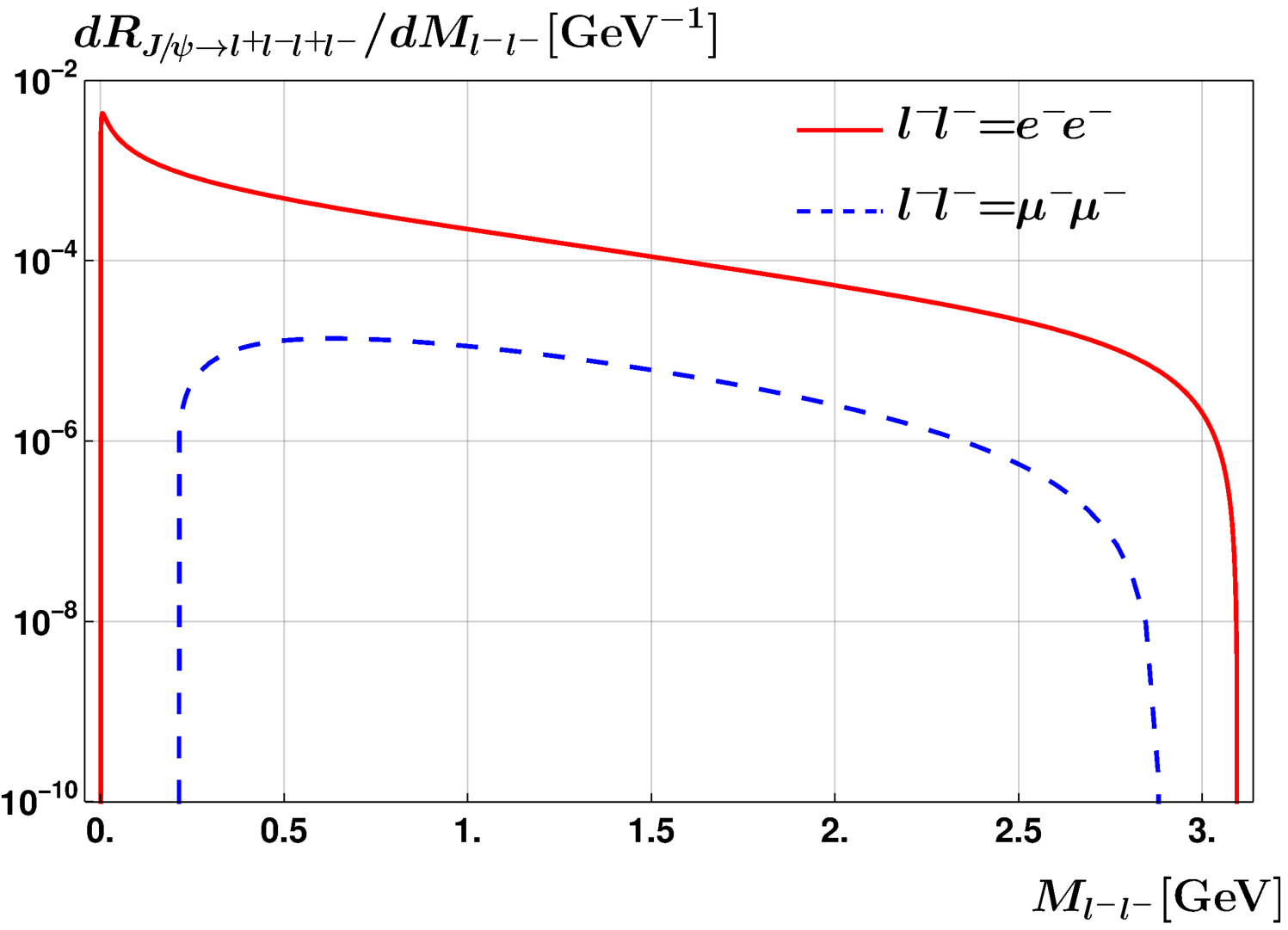}
  \includegraphics[width=0.45\textwidth]{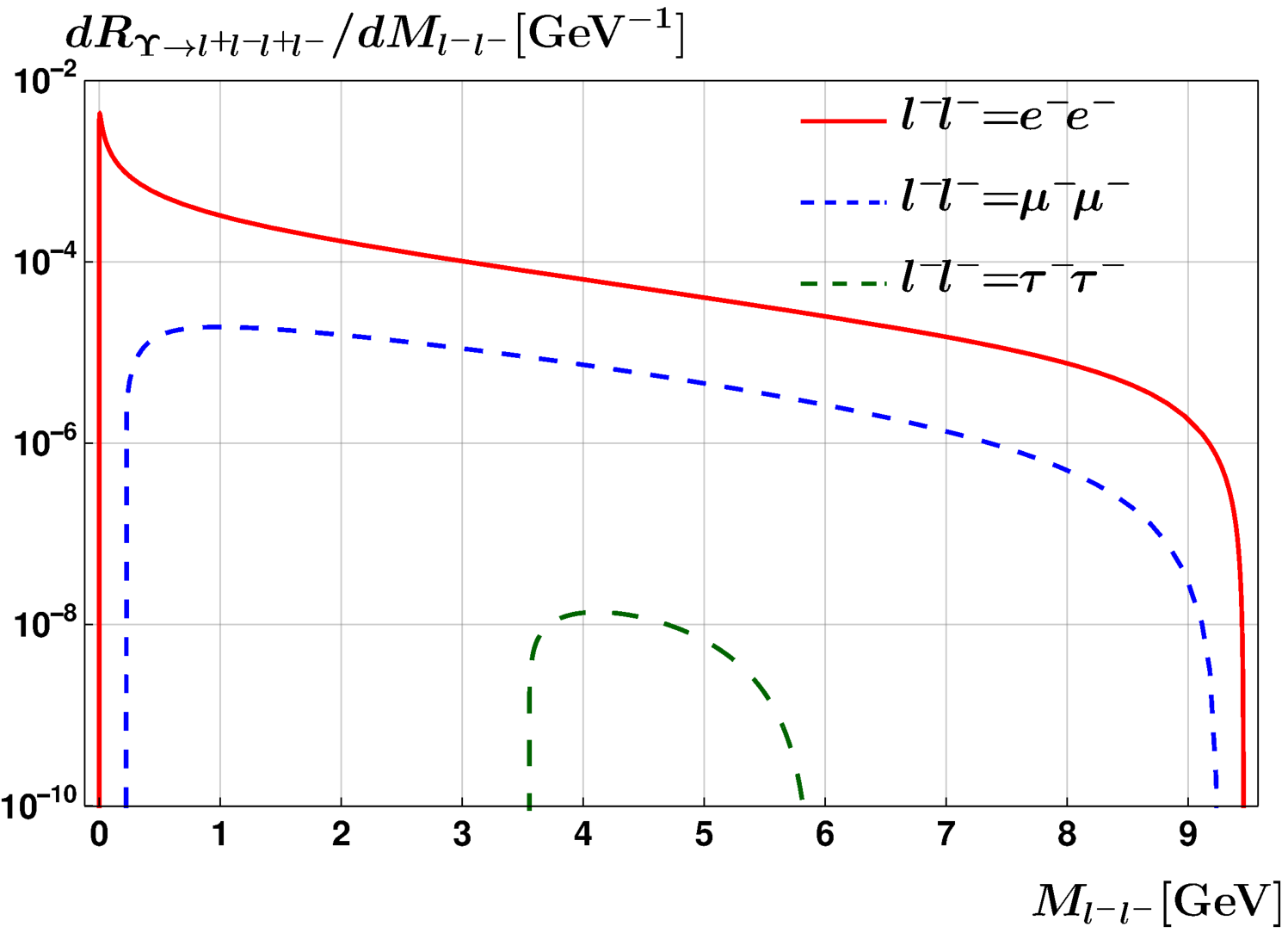}	
  \caption{ $e^-e^-$ ($\mu^-\mu^-$) invariant mass spectra in $V \to 4e(4\mu,4\tau)$, where $V$ stands for $\rho^0(\omega)$, $\phi$, $J/\psi$ and $\Upsilon$. }
	\label{ee:inv:spectrum:4e}
\end{figure}

\begin{figure}[hbtp]
	\centering
	\includegraphics[width=0.45\textwidth]{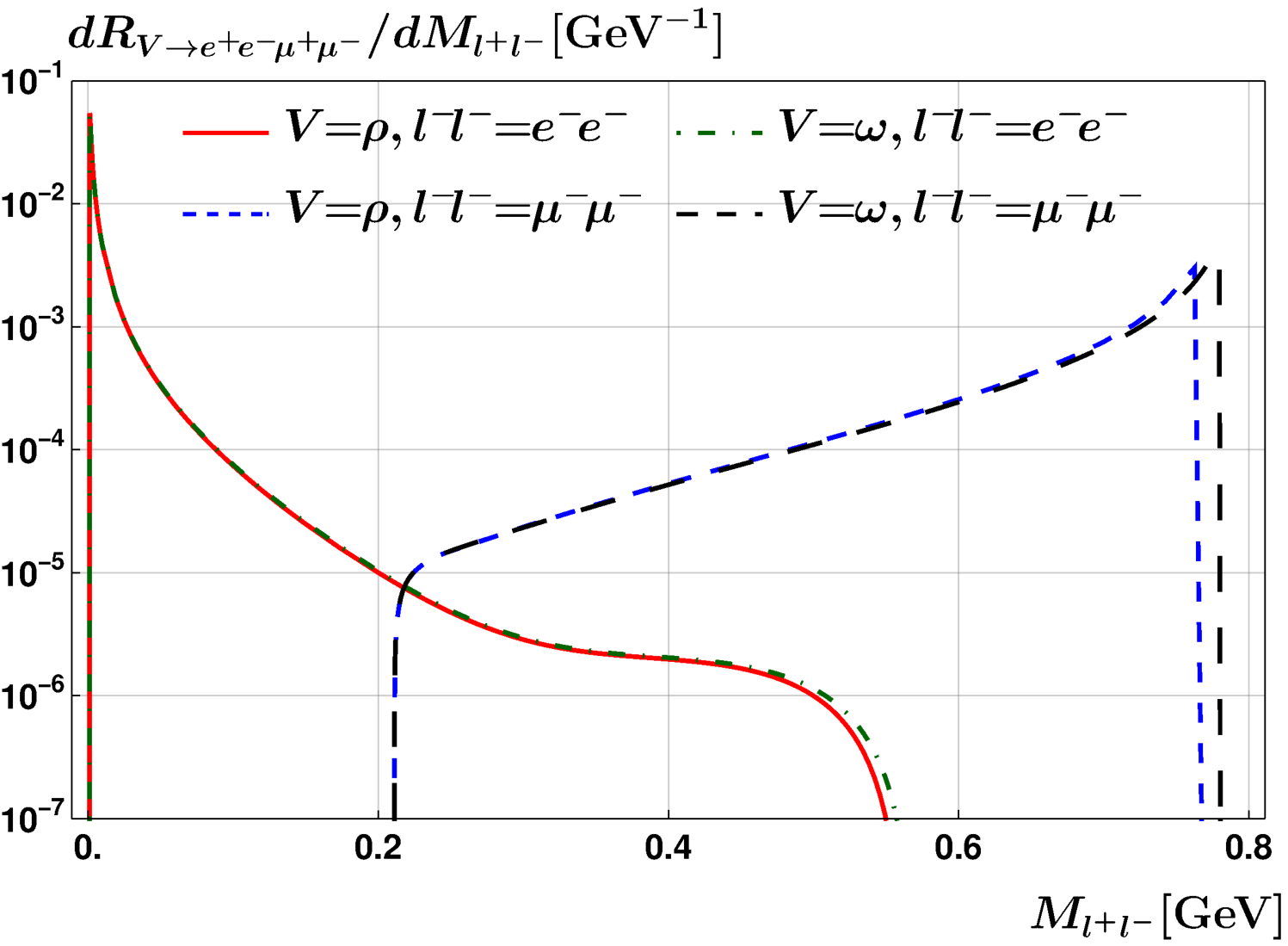}
  \includegraphics[width=0.45\textwidth]{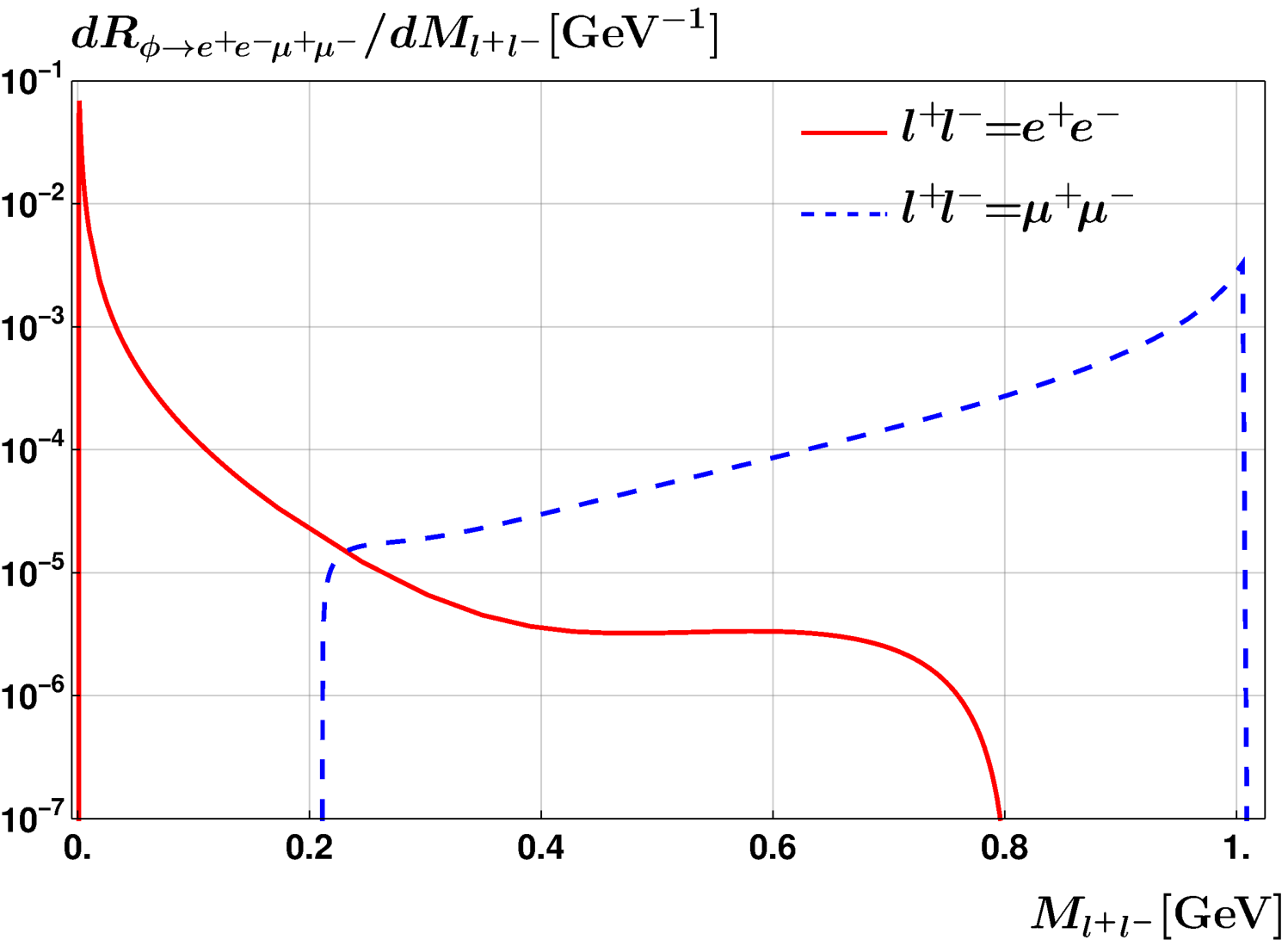}
  \includegraphics[width=0.45\textwidth]{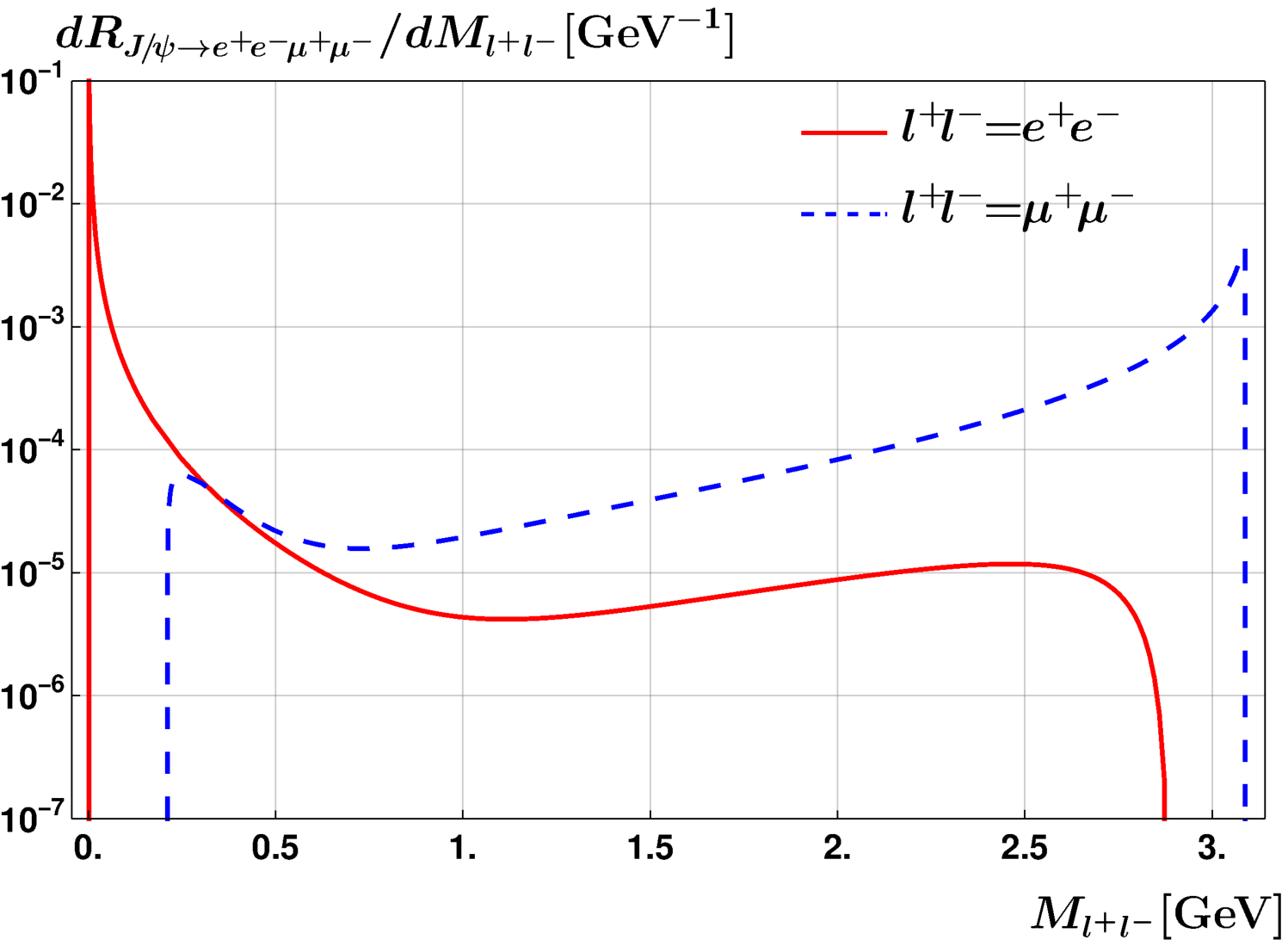}
	\includegraphics[width=0.45\textwidth]{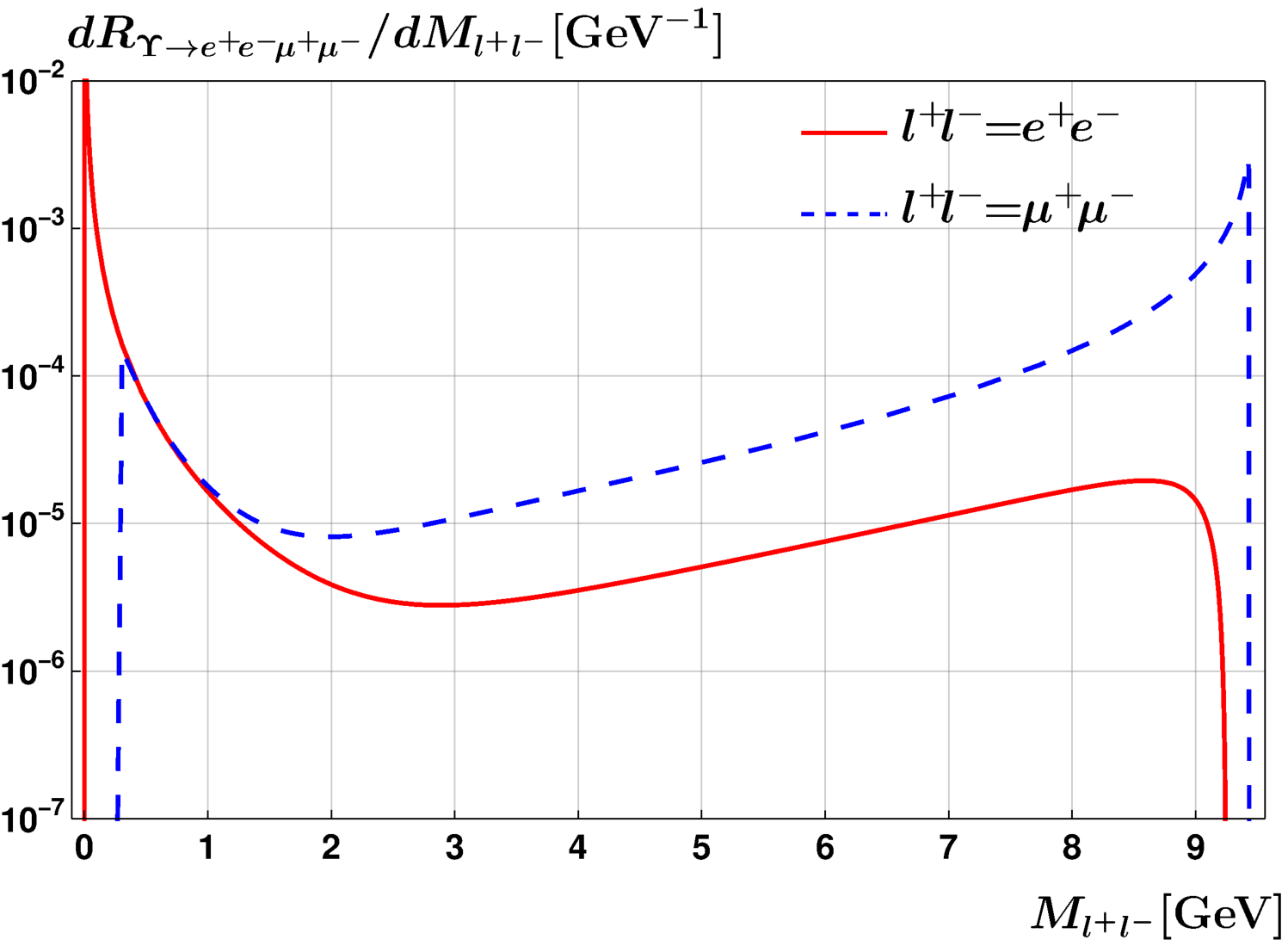}
	\caption{$e^+e^-$ ($\mu^+\mu^-$) invariant mass spectra in $V\to e^+e^- \mu^+\mu^-$, where $V$ stands for $\rho^0(\omega)$, $\phi$, $J/\psi$ and $\Upsilon$.}
	\label{e+e-:inv:spectrum:2e2mu}
\end{figure}

\begin{figure}[hbtp]
	\centering
	\includegraphics[width=0.45\textwidth]{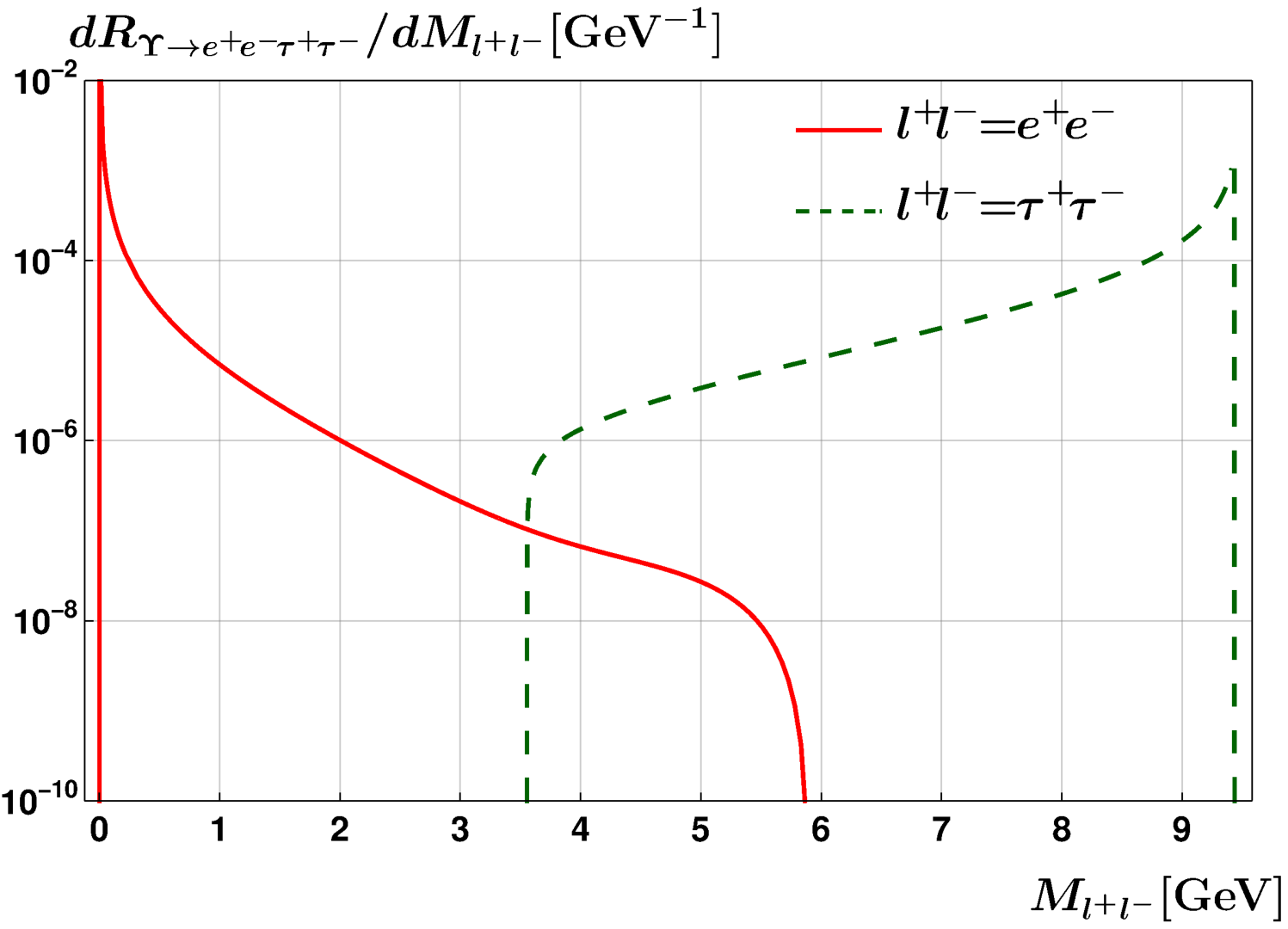}
	\includegraphics[width=0.45\textwidth]{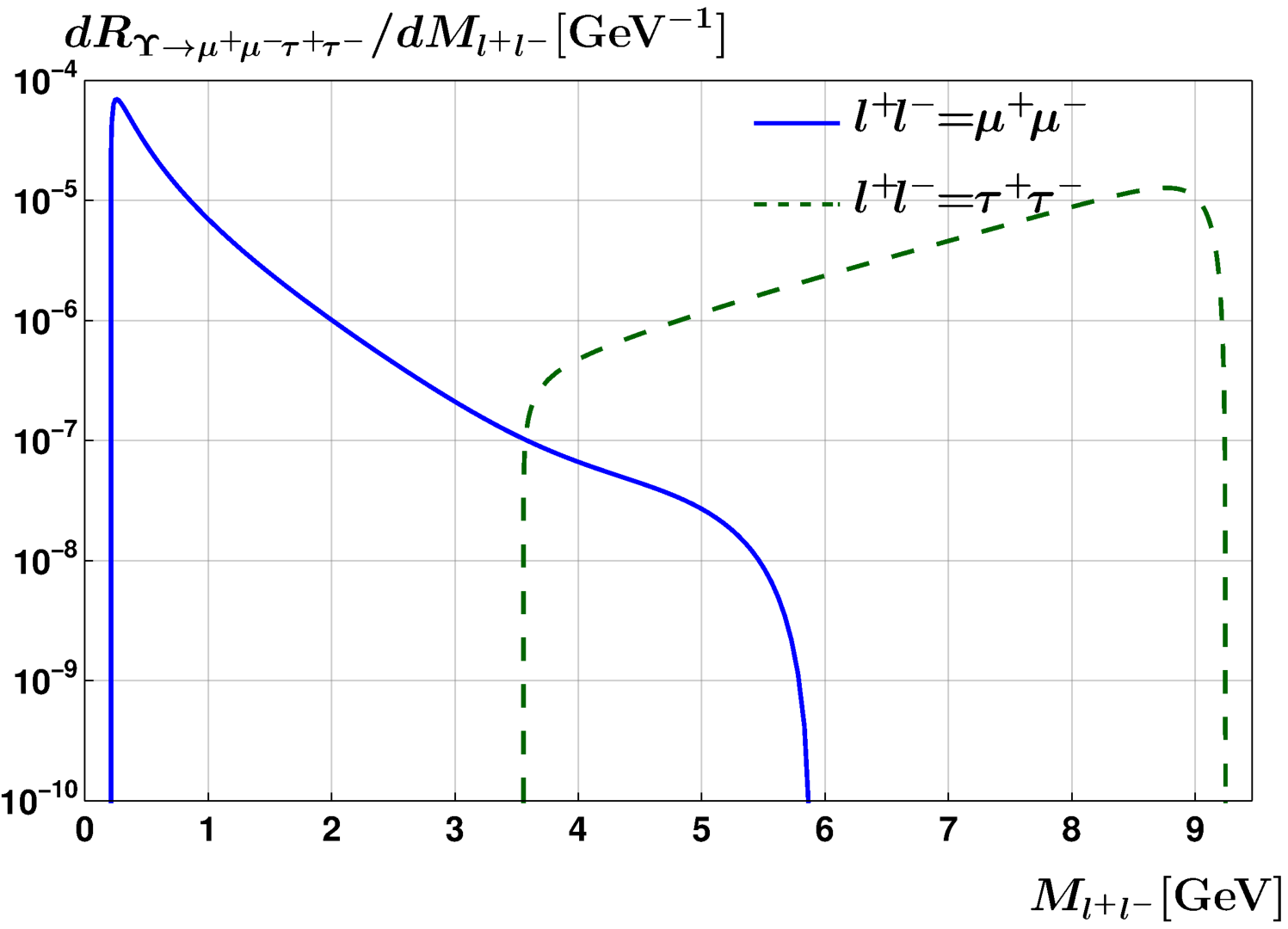}
	\caption{$e^+e^-$ ($\mu^+\mu^-,\tau^+\tau^-$) invariant mass spectra in $\Upsilon \to e^+e^-\tau^+\tau^-(\mu^+\mu^-\tau^+\tau^-)$.}
	\label{e+e-:inv:spectrum:2emu2tau}
\end{figure}

The four-body leptonic final state allows one to probe a variety of differential spectra, which
can sharpen the test of our QED predictions in a greater detail.
From Fig.~\ref{ee:inv:spectrum:4e} to Fig.~\ref{e+e-:inv:spectrum:2emu2tau},
we present the invariant mass spectra of various combination of lepton pairs in the final state.
It is desirable if the future experiments can confront our
predictions for invariant mass distributions.

In summary, in this work we have investigated a specific type of rare electromagnetic decays
of vector mesons, where the final state is composed of four charged leptons.
To the best of our knowledge,
the current work represents the first theoretical study of these processes.
For simplicity, we have only considered the lowest-order QED contribution,
which should already constitute decent predictions according to our experiences.
In sharp contrast to the predicted width for vector meson into a lepton pair,
where the lepton mass is usually discard, we must explicitly retain the finite lepton mass to regularize
the potential collinear/soft divergences. The branching fractions of $V\to 2(e^+e^-)$ is orders of magnitude 
greater than that of $V\to 2(\mu^+\mu^-)$, because of more pronounced triple-logarithmic enhancement received 
by the former channel.
We predict that the branching fractions associated with the
$J/\psi\to 2(e^+e^-), e^+e^-\mu^+\mu^-$ and
$\Upsilon\to 2(e^+e^-), e^+e^-\mu^+\mu^-, e^+e^-\tau^+\tau^-$ channels
all reach the order $10^{-5}$, which may indicate bright observation opportunity at {\tt BESIII}, {\tt Belle 2} and {\tt LHC} experiments
in near future.

\begin{acknowledgments}
We thank Haibo Li for interesting discussion that stimulate us to initiate this work.
We are also grateful to Xiaohui Liu for useful discussions on triple logarithms in the massless lepton
limit.
W.~C. wishes to acknowledge the warm hospitality extended to him by Theory Division of IHEP during his visit in COVID-19 pandemic.
The work of W.-C., Y.~J., Z.~M. and J.~P. is supported in part by the National Natural Science Foundation of China
under Grants No.~11925506, 11875263, No.~12070131001 (CRC110 by DFG and NSFC).
The work of X.-N. X. is supported in part by the National Natural Science Foundation of China under Grants No. 11905296.
\end{acknowledgments}

\appendix

\section{Four-body phase-space integration}\label{App:FourBodPhasSpacInt}

In this Appendix, we present some technical details about the four-body phase space integration.
To be specifical, let us consider $J/\psi(P)\to e^-(k_1)e^+(k_2)\mu^-(k_3)\mu^+(k_4)$.
The four-body phase-space integration can be carried out recursively, owing to its factorization property~\cite{pdg:2020}.
To our purpose, it is most convenient to lump the $e^+e^-$ into a compound particle,
and lump $\mu^+\mu^-$ into another compound particle.
The four-body phase space can then be factorized into the convolution of
three two-body phase space integration measures:
\begin{align}
  \int\!\! d\Pi_4 & \equiv \int \!\! \prod_{i=1}^4[dk_i](2\pi)^4\delta^{(4)}(P-\sum_{i=1}^4k_i)
\nn \\
&= \frac{1}{(2\pi)^2}\int \!\! dM_{12}^2dM_{34}^2d\Pi_{K_{12}K_{34}}d\Pi_{k_1k_2}d\Pi_{k_3k_4},
\label{4:body:phase:space:factorization}
\end{align}
where $d\Pi_{q_1q_2}\equiv[dq_1][dq_2](2\pi)^4\delta^{(4)}(Q-q_1-q_2)$ denotes the two-body phase-space integration measure $d\Pi_{2}$,
with $[dq]\equiv {d^3 q\over (2\pi)^3 2E_q}$.
In \eqref{4:body:phase:space:factorization}, we define the momentum of the compound system as
$K_{ab}\equiv k_a+k_b$, with its invariant mass $M_{ab}\equiv\sqrt{K_{ab}^2}$.

Since the two-body phase space $d\Pi_{2}$ is Lorentz invariant,
we can choose to calculate in any reference frame.
We prefer to work in the center-of-mass frame for each individual two-body phase space integral:
\bseq
\begin{align}
d\Pi_{K_{12}K_{34}} &= \frac{|\bm{K}_{12}|}{16\pi^2M_V}~d\cos\Theta d\Phi,
\\
d\Pi_{k_1k_2}       &= \frac{\sqrt{M_{12}^2-4m_e^2}}{32\pi^2M_{12}}~d\cos\theta_1d\varphi_1,
\\
d\Pi_{k_3k_4}       &= \frac{\sqrt{M_{34}^2-4m_\mu^2}}{32\pi^2M_{34}}~d\cos\theta_2d\varphi_2,
\end{align}
\eseq
where $M_V$ is the mass of the vector quarkonium, $\Theta$ and $\Phi$ denote the polar and azimuthal angles of $K_{12}$ in the rest frame of $J/\psi$,
$\theta_1$ and $\varphi_1$ denote the polar and azimuthal angles of $k_1$ in the rest frame of $K_{12}$,
and $\theta_2$ and $\varphi_2$ represent the polar and azimuthal angles of $k_3$ in the rest frame of $K_{34}$.
It is straightforward to obtain $|\bm{K}_{12}|=|\bm{K}_{34}|=\frac{1}{2M_V}\sqrt{(M_{12}^2+M_{34}^2-M_J^2)^2-4M_{12}^2M_{34}^2}$.
Due to rotational symmetry, the unpolarized squared amplitude is independent of $\Theta$ and $\Phi$,
and depends on the azimuthal angles $\varphi_1$ and $\varphi_2$ only through the difference $\varphi\equiv\varphi_1-\varphi_2$.
Integrating over dummy variables, and relabeling $\theta_1$ and $\theta_2$ by $\theta$ and $\theta^\prime$, we then have
\begin{align}
\notag \int d\Pi_4 = &\frac{1}{2^{11}\pi^6M_V}\int_{2m_e}^{M_V}\!\!dM_{12}\int_{2m_\mu}^{M_J-M_{12}}\!\!dM_{34}d\cos\theta d\cos\theta^\prime d\varphi\\
&\times|\bm{K}_{12}|\sqrt{(M_{12}^2-4m_e^2)(M_{34}^2-4m_\mu^2)}.
\end{align}

Because the squared amplitude is a Lorentz scalar, it can be evaluated in an arbitrary frame. We choose the rest frame of $J/\psi$
to compute the unpolarized decay rate.
We further assume $K_{12}$ and $K_{34}$ to move along the $z$ axis, while $k_3$ lies in the $y$-$z$ plane.
$k_2$ can be replaced by $K_{12}-k_1$, and $k_4$ can be replaced by $K_{34}-k_3$. The rest of momenta are parameterized by
\bseq
\begin{alignat}{2}
K_{12}^0    &= K_{34}^0=\sqrt{\bm{K}_{12}^2+M_{34}^2}, \quad &
\bm{K}_{12} &= -\bm{K}_{34}=(0,0,|\bm{K}_{12}|),
\\
k_1^\mu     &= L^\mu_{~\nu}(K_{12})k_1^{\prime\nu}, \quad &
k_3^\mu     &= L^\mu_{~\nu}(K_{34})k_3^{\prime\nu},
\end{alignat}
\eseq
with the reference momenta given by
\bseq
\begin{alignat}{2}
k_1^{\prime0}   &= \frac{M_{12}}{2}, \quad &
\bm{k}_1^\prime &= \frac{1}{2}\sqrt{M_{12}^2-4m_e^2}(\sin\theta\cos\varphi,\sin\theta\sin\varphi,\cos\theta),
\\
k_3^{\prime0}   &= \frac{M_{34}}{2}, \quad &
\bm{k}_3^\prime &= \frac{1}{2}\sqrt{M_{34}^2-4m_\mu^2}(\sin\theta^\prime,0,\cos\theta^\prime).
\end{alignat}
\eseq
Here $L^\mu_\nu(K_{ab})$ represents a Lorentz boost matrix (with $(a,b)= 1,2$ or $3,4$),
whose matrix elements are explicitly given by
\bseq
\begin{align}
L^0_{~0}(K_{ab}) &= \frac{K_{ab}^0}{M_{ab}},
\\
L^0_{~i}(K_{ab}) &= L^{i}_{~0} = \frac{1}{M_{ab}}K_{ab}^i,
\\
 L^i_{~j}(K_{ab}) &= \delta^i_{~j}+\frac{1}{M_{ab}\left(M_{ab}+K_{ab}^0\right)}K_{ab}^iK_{ab}^j,
\end{align}
\eseq
with the Latin indices $i,j=1,2,3$.


\begin{thebibliography}{30}

\bibitem{pdg:2020}
  P.A. Zyla et al. (Particle Data Group), Prog. Theor. Exp. Phys. 2020, 083C01 (2020).

\bibitem{Bodwin:1994jh}
  G.~T.~Bodwin, E.~Braaten and G.~P.~Lepage,
  Phys.\ Rev.\ D {\bf 51}, 1125 (1995)
  Erratum: [Phys.\ Rev.\ D {\bf 55}, 5853 (1997)].

\bibitem{Beneke:1997jm}
  M.~Beneke, A.~Signer and V.~A.~Smirnov,
  Phys. Rev. Lett. \textbf{80}, 2535-2538 (1998)
  doi:10.1103/PhysRevLett.80.2535
  [arXiv:hep-ph/9712302 [hep-ph]].

\bibitem{Asner:2008nq}
  D.~M.~Asner, T.~Barnes, J.~M.~Bian, I.~I.~Bigi, N.~Brambilla, I.~R.~Boyko, V.~Bytev, K.~T.~Chao, J.~Charles and H.~X.~Chen, \textit{et al.}
  Int. J. Mod. Phys. A \textbf{24}, S1-794 (2009)
  [arXiv:0809.1869 [hep-ex]].

\bibitem{Jia:2020csg}
  S.~Jia, X.~Zhou and C.~Shen,
  Front. Phys. (Beijing) \textbf{15}, no.6, 64301 (2020)
  [arXiv:2005.05892 [hep-ex]].

\bibitem{KLOE:2004uzx}
F.~Ambrosino \textit{et al.} [KLOE],
Phys. Lett. B \textbf{608}, 199-205 (2005)
[arXiv:hep-ex/0411082 [hep-ex]].

\bibitem{LHCb:2013itw}
R.~Aaij \textit{et al.} [LHCb],
JHEP \textbf{06}, 064 (2013)
[arXiv:1304.6977 [hep-ex]].

\bibitem{Hahn:2000kx}
  T.~Hahn,
  Comput.\ Phys.\ Commun.\  {\bf 140}, 418 (2001).

\bibitem{Shtabovenko:2020gxv}
  V.~Shtabovenko, R.~Mertig and F.~Orellana,
  Comput.\ Phys.\ Commun.\  {\bf 256}, 107478 (2020).

\bibitem{Anastasiou:2002yz}
  C.~Anastasiou and K.~Melnikov,
  Nucl.\ Phys.\ B {\bf 646}, 220 (2002).

\bibitem{Gehrmann-DeRidder:2003pne}
  A.~Gehrmann-De Ridder, T.~Gehrmann and G.~Heinrich,
  Nucl.\ Phys.\ B {\bf 682}, 265 (2004).

\bibitem{GehrmannDeRidder:2004tv}
A.~Gehrmann-De Ridder, T.~Gehrmann and E.~W.~N.~Glover,
Nucl. Phys. B \textbf{691}, 195-222 (2004)
doi:10.1016/j.nuclphysb.2004.05.017
[arXiv:hep-ph/0403057 [hep-ph]].

\bibitem{Catani:1999ss}
  S.~Catani and M.~Grazzini,
  Nucl.\ Phys.\ B {\bf 570}, 287 (2000).

\bibitem{Magnea:2018hab}
  L.~Magnea, E.~Maina, G.~Pelliccioli, C.~Signorile-Signorile, P.~Torrielli and S.~Uccirati,
  JHEP \textbf{12}, 107 (2018)
  [erratum: JHEP \textbf{06}, 013 (2019)].

\bibitem{Hahn:2016ktb}
  T.~Hahn,
  Comput.\ Phys.\ Commun.\  {\bf 207}, 341 (2016).

\end{thebibliography}
\end{document}